\documentclass[11pt,a4paper]{article}
\usepackage{jheppub}
\usepackage{subfigure}

\title{Descalarization by Quenching Charged Hairy Black Hole in asymptotically AdS spacetime}

\author[a]{Qian Chen,}

\author[a]{Zhuan Ning,}

\author[a,b]{Yu Tian,}

\author[c,d]{Bin Wang,}

\author[e]{Cheng-Yong Zhang}

\affiliation[a]{School of Physics, University of Chinese Academy of Sciences, Beijing 100049, China}

\affiliation[b]{Institute of Theoretical Physics, Chinese Academy of Sciences, Beijing 100190, China}

\affiliation[c]{Center for Gravitation and Cosmology, College of Physical Science and Technology, Yangzhou University, Yangzhou 225009, China}

\affiliation[d]{Shanghai Frontier Science Center for Gravitational Wave Physics, Shanghai Jiao Tong University, Shanghai 200240, China}

\affiliation[e]{Department of Physics and Siyuan Laboratory, Jinan University, Guangzhou 510632, China}

\emailAdd{chenqian192@mails.ucas.ac.cn}

\emailAdd{ningzhuan17@mails.ucas.ac.cn}

\emailAdd{ytian@ucas.ac.cn}

\emailAdd{wang\_b@sjtu.edu.cn}

\emailAdd{zhangcy@email.jnu.edu.cn}

\abstract{In this work, we study the real-time dynamics of the charged hairy black hole with the time-dependent source of scalar field in asymptotically anti-de Sitter (AdS) spacetime. 
The numerical results reveal a novel descalarization mechanism.
In order to obtain the hairy black hole as the initial data for the quench process, we first analyze the quasi-normal modes of the massive complex scalar field on the Reissner-Nordström anti-de Sitter (RN-AdS) black hole background.
We find the dominant unstable modes for large and small RN-AdS black holes come from the zero-damped modes and AdS modes, respectively. 
Then, the unstable RN-AdS black holes are perturbed to trigger the transition to hairy black holes.
With the hairy black hole in hand, we specify a time dependent scalar source for the system.
As the source is turned on, the electric charge, energy and scalar condensation of the system start to oscillate with the entropy increasing monotonically.
Finally, with the decay of the scalar source, the system gradually settles down to a new state.
Interestingly, the final state of the evolution could be a hairy black hole with less scalar condensation, a RN-AdS black hole or a Schwarzschild-AdS black hole, which depends on the quench strength.
However, as long as the quench strength is large enough, the system always loses all the electric charge and converges to the Schwarzschild-AdS black hole.}


\arxivnumber{}

\begin{document}
	
\maketitle

\section{Introduction}\label{sec:In}
In general relativity, a stationary black hole in electrovacuum asymptotically flat spacetime can be totally characterized by its mass, angular momentum and charge \cite{Chrusciel:2012jk}. However, in the presence of non-linear matter sources or in many alternative gravity theories, a variety of black hole solutions owing further characters, which are collectively referred to as hairy black holes, have been found \cite{Bizon:1994dh,Volkov:1998cc,Herdeiro:2015waa}. 
Among these, the black holes with scalar hair have attracted much attention in the last few decades. Especially, it was found that in asymptotically flat spacetime, Kerr black holes suffer superradiant instability under massive free complex scalar field perturbations and develop into hairy black holes \cite{Herdeiro:2014goa,Herdeiro:2015gia}. These hairy generalisations of the Kerr solution bifurcate from the zero modes of the superradiant instability, which often indicate the existence of superradiant bound states  \cite{Brito:2015oca}. 
Akin to the rotational superradiance of Kerr black hole, the charged superradiance happens to    Reissner-Nordstr\"om (RN)  black holes \cite{Bekenstein:1973mi}. However, the RN black hole can not support the complex scalar hair since no zero modes or bound states exist in asymptotically flat spacetime. The situation is changed when the charged black hole is placed in a cavity \cite{Sanchis-Gual:2015lje,Sanchis-Gual:2016tcm} or in an asymptotically anti-de Sitter (AdS) spacetime \cite{Bosch:2016vcp,Dias:2016pma}, in which the RN black hole becomes superradiantly unstable and develops into a stable hairy black hole.

The superradiant instability of RN black hole requires confinement due to the AdS boundary or a reflecting mirror enclosing the black hole \cite{Basu:2010uz,Dias:2011tj}. 
On the other hand, besides the superradiant instability, the extremal RN-AdS black holes further have the near-horizon condensation instability when the effective mass of the scalar perturbation is smaller than the Breitenlohner-Freedman (BF) bound of the near-horizon AdS$_2$ geometry \cite{Breitenlohner:1982jf}. 
The near horizon instability has been studied extensively in the context of gauge-gravity duality such as holographic superconductors and superfluids \cite{Gubser:2008px,Hartnoll:2008vx,Herzog:2009xv}, and is sometimes referred to as the IR instability. The near horizon geometry of the near extremal black hole gives a collection of zero-damped modes whose frequencies cluster onto the real axis in the extremal limit  
\cite{Yang:2012pj,Zimmerman:2015trm,Zimmerman:2016qtn}. The interplay between the superradiant instability and near horizon instability of the RN-AdS black hole has been studied in \cite{Dias:2016pma,Bosch:2019anc}. In this work, we will disclose further the dominant unstable modes for large and small RN-AdS black holes. We find that IR instability is suppressed in small RN-AdS black hole, while the superradiant instability plays a role in both large and small RN-AdS black holes.

The near horizon instability is a kind of tachyonic instability, which has prompted a flurry of activity on the black hole spontaneous scalarization recently \cite{Doneva:2017bvd,Silva:2017uqg,Antoniou:2017acq,Cunha:2019dwb,Dima:2020yac,Herdeiro:2020wei,Berti:2020kgk,Herdeiro:2018wub,Zhang:2021nnn}. The nonlinear studies on the time evolution of the black holes  confirmed that the resulting  black holes are scalarized  \cite{Ripley:2020vpk,Doneva:2021dqn,East:2021bqk,Zhang:2021etr,Kuan:2021lol,Xiong:2022ozw,Luo:2022roz}. On the other hand, the scalar hair of the hairy black hole can be removed through some dynamical processes. 
The first mechanism for descalarization was realized in the binary black hole merger \cite{Silva:2020omi,Doneva:2022byd,Elley:2022ept}. However, these works were limited in the decoupling limit where the nonlinear scalar equation of motion was only evolved on the background from general relativity. Then another mechanism for descalarization by accretion was introduced in \cite{Corelli:2021ikv,Kuan:2022oxs,Zhang:2022cmu,Liu:2022eri,Liu:2022fxy,Niu:2022zlf}. These works considered the full nonlinear evolution of black holes under large perturbation. The scalar hair exists only when the charge to mass ratio and coupling strength are large for charged black holes in Einstein-Maxwell-scalar theories, or when the mass to the Gauss-Bonnet coupling constant ratio is small for Schwarzschild black holes in scalar-Gauss-Bonnet theories. In both theories, the dynamical descalarization occurs when these ratios are changed to the parameter region where only bald black holes survive. 

In these previous works, the dynamics were studied in the microcanonical ensemble since the total mass and charge are fixed during the evolution. In this work, we introduce a new mechanism for the descalarization by quenching a charged hairy black hole, which is the endpoint of a superradiant unstable RN-AdS black hole under complex scalar perturbation \cite{Bosch:2016vcp,Bosch:2019anc}. We put the charged hairy AdS black hole in an open environment. By turning on a time dependent source of the complex scalar field at the AdS boundary, we inject energy and charge into the bulk. This process decreases the charge to mass ratio such that the scalar hair of the initial hairy black hole can be removed. The endpoint of this process can be a hairy black hole with less scalar hair, a RN-AdS black hole or a Schwarzschild-AdS black hole. If the quench strength is large enough, the system always loses all the electric charge and scalar hair, and converges to the Schwarzschild-AdS black hole.

The organization of the paper is as follows. In Section \ref{sec:Md} we introduce the model we considered. In Section \ref{sec:instability}, we analyse the instabilities and the quasinormal modes of the RN-AdS black hole under complex scalar perturbation. In Section \ref{sec:nonlinear} we investigate the nonlinear dynamics of the scalarization and descalarization. Finally in the last section, we will summarize the obtained results.

\section{Model}\label{sec:Md}
We consider the Einstein-Maxwell-Scalar system in the four-dimensional asymptotically anti-de Sitter spacetime with the Lagrangian density
\begin{equation}
	\mathcal{L}=R-2\Lambda-\frac{1}{4}F_{\mu\nu}F^{\mu\nu}-D_{\mu}\psi (D^{\mu}\psi)^*-m^{2}|\psi|^{2},
	\label{eq:Lagrangian}
\end{equation}
where $D_{\mu}=\nabla_{\mu}-iqA_{\mu}$ is the gauge covariant derivative and the negative cosmological constant is chosen to be $\Lambda=-3$ by setting the AdS scale $L=1$.
The mass of the scalar field should respect the Breitenlohner-Freedman (BF) bound $-9/4\leq m^{2}L^{2}\leq 0$ to make the theory stable \cite{Breitenlohner:1982jf}.
In this case, the scalar field has two branches of the convergent asymptotic solutions near the boundary
\begin{equation}
	\psi\sim \psi_{(-)}r^{-(3-\Delta)}+\psi_{(+)}r^{-\Delta},\quad r\rightarrow \infty,
	\label{eq:asy_solutions}
\end{equation}
where the conformal dimension of the scalar operator satisfies $\Delta(\Delta-3)=m^{2}L^{2}$, and its discriminant gives rise to the BF bound. 
The coefficients $\psi_-$ and $\psi_+$ are the source and response, respectively. 
In what follows, we will take $m^{2}L^{2}=-2$ for definiteness to make $\Delta=2$.

In this model, the Einstein equation is
\begin{equation}
	G_{\mu\nu}=-\Lambda g_{\mu\nu}+T^{\psi}_{\mu\nu}+T^{M}_{\mu\nu},
	\label{eq:Einstein_equation}
\end{equation}
where the stress-energy tensors of the gauge field and scalar field are
\begin{subequations}
	\begin{align}
		T^{M}_{\mu\nu}&=\frac{1}{2}g^{\rho\sigma}F_{\mu\rho}F_{\nu\sigma}-\frac{1}{8}F_{\alpha\beta}F^{\alpha\beta}g_{\mu\nu},\\
		T^{\psi}_{\mu\nu}&=\frac{1}{2}\left[D_{\mu}\psi (D_{\nu}\psi)^*+D_{\nu}\psi (D_{\mu}\psi)^*\right]-\frac{1}{2}\left[D_{\alpha}\psi (D^{\alpha}\psi)^*+m^{2}\psi{\psi}^*\right]g_{\mu\nu}.
	\end{align}
\end{subequations}
The equations of motion for the gauge field and scalar field are
\begin{subequations}
	\begin{align}
		\nabla_{\nu}F^{\nu\mu}&=qJ^{\mu},\label{eq:Mx}\\
		D^{\mu}D_{\mu}\psi&=m^{2}\psi,\label{eq:KG}
	\end{align}
\end{subequations}
where $J^{\mu}=i\left[{\psi}^* D^{\mu}\psi-\psi (D^{\mu}\psi)^*\right]$ is the Noether current of the complex scalar.

The RN-AdS black hole solves the field equations with metric
\begin{equation}
	ds^{2}=-f(r)dt^{2}+\frac{dr^{2}}{f(r)}+r^{2}d\Omega^{2}_{2},
\end{equation}
where
\begin{equation}
	f(r)=1-\frac{2M}{r}+\frac{Q^{2}}{4r^{2}}+\frac{r^{2}}{L^{2}},
\end{equation}
and Maxwell potential
\begin{equation}
	A_{\mu}dx^{\mu}=\left(\mu-\frac{Q}{r}\right)dt,
\end{equation}
and with the vanishing scalar field. 
The parameters $M$ and $Q$ are the mass and the electrical charge of the black hole, respectively.
The constant $\mu$ is a pure gauge, which is set to $0$ in most of this paper. However, to study the near-horizon geometry of the RN-AdS black hole in Sec \ref{subsec:IR}, we will take $\mu=Q/r_{+}$ for convenience, where $r_{+}$ is the outer horizon radius.
The Hawking temperature is given by
\begin{equation}
	T=\frac{1}{4\pi}f'(r_{+})=\frac{1}{4\pi r_{+}}\left(1-\frac{Q^{2}}{4r_{+}^{2}}+\frac{3r^{2}_{+}}{L^{2}}\right).
\end{equation}
Due to the positive definiteness of the Hawking temperature, we have
\begin{equation}
	Q\leqslant 2r_{+}\sqrt{1+\frac{3r^{2}_{+}}{L^{2}}}=Q_{c}.
\end{equation}

On the other hand, the RN-AdS black hole can be described  in terms of the inner and outer horizon radius, $r_{-}$ and $r_{+}$, which are the real roots of the equation $f(r)=0$.
In this case, the metric function is expressed by 
\begin{equation}
	f(r)=L^{-2}(r-r_{+})(r-r_{-})[1+(r_{+}+r_{-})r^{-1}+(r^{2}_{+}+r^{2}_{-}+r_{+}r_{-}+L^{2})r^{-2}],
\end{equation}
from which we can get the mass and the electrical charge of the black hole immediately
\begin{subequations}
	\begin{align}
		M&=\frac{1}{2}L^{-2}(r_{+}+r_{-})(r^{2}_{+}+r^{2}_{-}+L^{2}),\\
		Q^{2}&=4L^{-2}r_{+}r_{-}(r^{2}_{+}+r^{2}_{-}+r_{+}r_{-}+L^{2}).
	\end{align}
\end{subequations}

\section{Instability} \label{sec:instability}
In this section, we take the complex scalar field with mass $m$ and charge $q$ as a probe field in the RN-AdS background.
For the linear analysis, we take the Klein-Gordon equation (\ref{eq:KG}) at the linear order
\begin{equation}
	0=r^{-2}\left(r^{2}f \delta\psi'  \right)'- f^{-1}\partial^{2}_{t} \delta\psi  + 2 i q {A_{t}}f^{-1}\partial_{t} \delta\psi +\left( q^{2} {A^{2}_{t}} f^{-1} - m^{2}  \right)\delta\psi,\label{eq:KG_linear}
\end{equation}
where the prime stands for the derivative with respect to the radius $r$.
The linear perturbations of the metric field and Maxwell potential have no contribution for the linear order Klein-Gordon equation due to the vanishing scalar field in the background.
We will analyze this equation in various situations.

We begin with the IR instability and the superradiant instability of RN-AdS black hole in Subsec \ref{subsec:IR} and in Subsec \ref{subsec:Sup} respectively, which can trigger the transition from a RN-AdS black hole to a hairy black hole.
Then, we describe the continued fraction method and persent the numerical results of the quasinormal frequencies in Subsec \ref{subsec:pertur}.

\subsection{IR instability }\label{subsec:IR}
We have set the mass of the scalar field to obey the BF bound in AdS$_{4}$, which ensure the asymptotically AdS$_{4}$ solutions are stable in the UV region. 
But so far, the behavior of the scalar field in the interior of the spacetime is free.
In particular, the structure of the spacetime in the IR region may impose an additional constraint on the behavior of the scalar field.
For an extremal RN-AdS black hole with spherical horizon topology, the near-horizon geometry is the direct product of AdS$_{2}$ with a sphere $S^{2}$.
So we expect that the system is driven to a hairy black hole with the near horizon scalar condensation when the effective mass of the scalar field at the horizon violates the BF bound in AdS$_{2}$.
Holographically, the scalar condensation signals a second order phase transition from RN-AdS black hole to hairy black hole, which is dual to a superconducting phase.

To consider the near-horizon geometry of an extremal RN-AdS black hole, we introduce the new coordinates
\begin{equation}
	t=L^{2}_{\text{AdS}_{2}}\lambda^{-1}\widetilde{t},\quad r=r_{+}+\lambda\widetilde{r},
\end{equation}
and let $\lambda\rightarrow 0$.
Under this limit and taking
\begin{equation}
	L^{-2}_{\text{AdS}_{2}}=6L^{-2}+r^{-2}_{+},
\end{equation}
we can get the near-horizon solution  with metric
\begin{equation}
	ds^{2}_{h}=L^{2}_{\text{AdS}_{2}}\left(-\widetilde{r}^{2}d\widetilde{t}^{2}+\frac{d\widetilde{r}^{2}}{\widetilde{r}^{2}}\right)+r^{2}_{+}d\Omega^{2}_{2},
\end{equation}
and Maxwell potential
\begin{equation}
	A_{a}dx^{a}=L^{2}_{\text{AdS}_{2}}Q_{c}r_{+}^{-2} \widetilde{r}d\widetilde{t}\equiv\widetilde{Q}_{c}\widetilde{r}d\widetilde{t}.
\end{equation}
Here $\widetilde{Q}_{c}=L^{2}_{\text{AdS}_{2}}Q_{c}r_{+}^{-2}$. Applying this near-horizon limit and taking the linear perturbation of the scalar field as $\delta\psi=e^{-iwt}\Psi_0$, the equation (\ref{eq:KG_linear}) reduces to the linear order Klein-Gordon equation for a scalar field around an AdS$_{2}$ background
	\begin{equation}
	0=\partial_{\widetilde{r}}(\widetilde{r}^{2}  \partial_{\widetilde{r}}\Psi_0) + \left[\left( w +q\widetilde{Q}_{c} \widetilde{r}\right)^{2}\widetilde{r}^{-2} - m^{2}L^{2}_{AdS_{2}}\right]\Psi_0(r),
\end{equation}
with a Maxwell potential $A_{\widetilde{t}}=\widetilde{Q}_{c}\widetilde{r}$. The above equation gives the asymptotic behavior of the scalar field near the horizon.
\begin{equation}
	\Psi_0\sim\phi_{-}\widetilde{r}^{-\widetilde{\Delta}_{-}}+\phi_{+}\widetilde{r}^{-\widetilde{\Delta}_{+}},\quad \text{with }\widetilde{\Delta}_{\pm}=\frac{1}{2}\pm\frac{1}{2}\sqrt{1+4m^{2}_{\text{eff}}L^{2}_{\text{AdS}_{2}}}
\end{equation}
where
\begin{equation}
	m^{2}_{\text{eff}}L^{2}_{\text{AdS}_{2}}=m^{2}L^{2}_{\text{AdS}_{2}}-q^{2}\widetilde{Q}_{c}^{2} \label{eq:mass}
\end{equation}
The IR instability occurs when the effective mass (\ref{eq:mass}) of the scalar field near horizon violates the AdS$_{2}$ BF bound
\begin{equation}
	m^{2}_{\text{eff}}L^{2}_{\text{AdS}_{2}}\leq-\frac{1}{4}.
\end{equation}
Thus, it shows that an extremal RN-AdS is unstable if the charge of the scalar field obeys
\begin{equation}
	4q^{2}L^{2}\geq \left[m^{2}L^{2}+\frac{3}{2}+\frac{1}{4}(r_{+}/L)^{-2}\right]\left[\frac{6+(r_{+}/L)^{-2}}{3+(r_{+}/L)^{-2}}\right]
\end{equation}
Taking the large RN-AdS limit $r_{+}/L\rightarrow\infty$, we have
\begin{equation}
	4q^{2}L^{2}\geq  2\left(m^{2}L^{2}+\frac{3}{2}\right)-\frac{1}{3}m^{2}L^{2}(r_{+}/L)^{-2}+O((r_{+}/L)^{-4}).
\end{equation}
This indicates a large RN-AdS black hole with sufficiently large $q$ suffers from the IR instability.
However, under the opposite limit $r_{+}/L\rightarrow 0$, it is not possible to trigger the IR instability since the AdS$_{2}$ BF bound is hard to be violated
\begin{equation}
	4q^{2}L^{2}\geq \frac{1}{4}(r_{+}/L)^{-2}+O(1).
\end{equation}

We have expounded that the IR instability occurs to an extremal RN-AdS black hole in which the effective mass of the scalar field near the  horizon violates the BF bound in AdS$_{2}$.
Due to the continuity of geometrical configuration, this instability is expected to extend to near-extremal configuration.
Moreover, we show that the IR instability is more likely to happen to the large RN-AdS black hole.
Particularly, there is a near horizon scalar condensation mechanism for the planar RN-AdS black hole, which can be regarded as the spherical RN-AdS black hole in the limit $r_{+}/L\rightarrow\infty$.
However, the IR instability is suppressed for small RN-AdS black hole, which is different from the superradiant instability, as we will discuss in the next subsection.

\subsection{Superradiance condition}\label{subsec:Sup}
Superradiance is one of the most famous mechanisms that destabilize black holes.
Under superradiant scattering, waves with suitabe tuned oscillating frequency scattering off the black hole can be amplified.
If the waves are confined, the outgoing waves can be reflected back inwards and interact with the black hole repeatedly, such that the black hole keeps converting energy to the wave and the wave amplitude grows exponentially, which drives the system to become unstable.
We will derive the superradiance condition for RN-AdS black hole from the process of superradiant scattering.

We consider a mode solution of massive scalar field $\delta\psi=r^{-1}e^{-iwt}\Psi_1$ with real frequency $w$ in the RN-AdS background.
The linear order Klein-Gordon equation (\ref{eq:KG_linear}) can be reduced to the ordinary differential equation
\begin{equation}
	0=f\left(f \Psi'_1  \right)' + \left[\left(\widetilde{w}-qQr^{-1}\right)^{2}  - m^{2}f -r^{-1}f'f \right]\Psi_1,
\end{equation}
where $\widetilde{w}=w+q\mu$. 
After using a new variable $dr_{*}=f^{-1}dr$, called the tortoise coordinate, the above equation has the Schr$\ddot{\text{o}}$dinger-like form
\begin{equation}
	0=\frac{d^{2}}{dr^{2}_{*}}\Psi_1 + V(r,\widetilde{\omega})\Psi_1.\label{eq:super}
\end{equation}
Here $V(r,\widetilde{\omega})$ is the effective potential. 
Considering a unit amplitude incident wave coming from the AdS boundary and scattering off the potential with reflection and transmission amplitudes $\mathcal{A}_{\infty}$ and $\mathcal{A}_{_{+}}$ respectively, we can get the follow asymptotic solutions of the equation (\ref{eq:super})
\begin{equation}
	\Psi_1\sim
	\begin{cases}
		\mathcal{A}_{_{+}}\text{exp}[-i(\widetilde{w}-qQr_{+}^{-1})r_{*}],&r\rightarrow r_{+},\\
		\text{exp}\left(-i\widetilde{w} r_{*}\right)+\mathcal{A}_{\infty}\text{exp}\left( i\widetilde{w} r_{*}\right),&r\rightarrow\infty.
	\end{cases}
\end{equation}
Due to the real effective potential $V(r,\widetilde{\omega})$, it is easy to prove if a function $\Psi_1$ solves the equation (\ref{eq:super}) and then its complex conjugate function ${\Psi}^*_1$ is another linearly independent solution.
The Wronskian determinant between the two independent solutions $\Psi_1$ and ${\Psi}^*_1$
\begin{equation}
	W=\Psi_1\frac{d}{dr_{*}} {\Psi}^*_1-{\Psi}^*_1\frac{d}{dr_{*}}\Psi_1,\label{eq:Wronskian}
\end{equation}
is conserved with respect to the variable $r_{*}$.
Evaluating the determinant (\ref{eq:Wronskian}) at the horizon and AdS boundary respectively, we have
\begin{equation}
	1-|\mathcal{A}_{\infty}|^{2}=|\mathcal{A}_{_{+}}|^{2}(\widetilde{w}-qQr_{+}^{-1})\widetilde{w}^{-1}.
\end{equation}
The   amplification of the reflection wave occurs if   following condition is satisfied: 
\begin{equation}
	0<\widetilde{w}<qQr_{+}^{-1}.\label{eq:surcon}
\end{equation}

We will show the superradiance condition (\ref{eq:surcon}) is easy to satisfy for small RN-AdS black hole with sufficiently large $q$.
In the limit $r_{+}/L\rightarrow 0$, the effects of the black hole may be neglected due to $M,Q\sim r_{+}\ll L$.
The linear order Klein-Gordon equation (\ref{eq:KG_linear}) reduces to 
\begin{equation}
	0=(1+r^{2}L^{-2}) \Psi''+(4rL^{-2}  +2r^{-1}) \Psi'  + \left[\widetilde{w}^{2} (1+r^{2}L^{-2})^{-1} - m^{2}  \right]\Psi,\label{eq:perads}
\end{equation}
which describes a perturbation $\delta\psi=e^{-iwt}\Psi$ of scalar field in pure AdS with a constant Maxwell potential $A_{t}=\mu$.
By defining a new variable and a new radial function
\begin{equation}
	x=1+r^{2}L^{-2},\quad \Psi=x^{\widetilde{w}L/2}F(x),
\end{equation}
one can find the radial function $F(x)$ solves the hypergeometric equation
\begin{equation}
		0=x(1-x)\frac{d^{2}}{dx^{2}}F(x)+\left[\gamma-\left(\alpha+\beta+1\right)x\right]\frac{d}{dx}F(x)-\alpha\beta F(x),  \\ 
\end{equation}
where
\begin{equation}
	\alpha=\frac{1}{2}\left(\widetilde{w}L+\Delta\right),\quad \beta=\frac{1}{2}\left(\widetilde{w}L-\Delta+3\right),\quad\gamma=\widetilde{w}L+1,
\end{equation}
and $\Delta$ is defined in (\ref{eq:asy_solutions}).
After imposing the no-source boundary condition, namely $\psi_{(-)}=0$ in (\ref{eq:asy_solutions}), the equation (\ref{eq:perads}) has the normalizable solutions
\begin{equation}
	\Psi_{n}=\mathcal{A} x^{-\frac{1}{2}\Delta}F\left(\frac{1}{2}\left(\Delta+\widetilde{w}_{n}L\right),\frac{1}{2}\left(\Delta-\widetilde{w}_{n}L\right);\Delta-\frac{1}{2};x^{-1}\right),
\end{equation}
with the frequency spectrum
\begin{equation}
	\widetilde{w}_{n}L=\Delta+2n,
\end{equation}
where $\mathcal{A}$ is an arbitrary amplitude and $n$ is a non-negative integer.
Thus, there is a set of normal modes of a   complex scalar field in the limit $r_{+}/L\rightarrow 0$, which   satisfy the superradiance condition
\begin{equation}
	0<\Delta+2n<qQr_{+}^{-1}L,
\end{equation}
with sufficiently large $q$.

In the previous subsection, we have seen that a large RN-AdS black hole is unstable to the IR instability.
However, it is suppressed for small RN-AdS black holes. 
In the present subsection, for the small RN-AdS black hole limit, we have found there may exist a series of modes satisfying the superradiance condition.
We therefore expect these modes to be unstable.
It has been proved in \cite{Dias:2016pma} using matched asymptotic expansions.
At high order the frequency will have a positive imaginary part if the mode satisfies the superradiance condition, which signals exponential growth in nonlinear evolution.

\subsection{Quasi-normal mode}\label{subsec:pertur}
The IR instability and superradaint instability of RN-AdS black hole have been found analytically in the limit $r_{+}/L\rightarrow\infty$ and $r_{+}/L\rightarrow 0$, respectively.
In the present subsection, we will further investigate these two instability numerically to reveal their behavior away from the limit. 

There are numerous methods to calculate the quasi-normal modes of the black hole perturbation, such as the WKB method, the continued fraction method, the asymptotic iteration method, the generalized eigenvalue method and so on \cite{Konoplya:2011qq}.
In this work, we use the continued fraction method to solve the eigenvalue problem of the perturbation equation.
Comparing with other numerical method, the continued fraction method is considered to be the most accurate one \cite{Zhang:2015jda}.
We will begin with the description of the continued fraction method and then give the numerical results for RN-AdS black hole in various situations.

In spherically symmetric background, we consider a mode of the probe scalar field with $\delta\psi= {e^{-iwt}\Phi}$. The linear order Klein-Gordon equation (\ref{eq:KG_linear}) has the form
\begin{equation}
	0=r^{-2}\left(r^{2}f \Phi'  \right)'+ \left[\left(\widetilde{w}-qQr^{-1}\right)^{2} f^{-1} - m^{2}  \right]\Phi.\label{eq:CFM_KG}
\end{equation}
Note that now we consider the system as an eigenvalue problem and $\widetilde{w}=w+q\mu$ is complex. After imposing the ingoing condition at the horizon and the no-source condition at the AdS boundary, the asymptotic solution of above equation is
\begin{equation}
	\Phi\sim
	\begin{cases}
		(r-r_{+})^{N_{+}},&r\rightarrow r_{+},\\
		r^{\Delta},&r\rightarrow\infty,
	\end{cases}
\end{equation}
where
\begin{equation}
	N_{+}=- i L^{2}r_{+}\left(r_{+}\widetilde{w}-qQ\right)[(3r^{2}_{+}+2r_{+}r_{-}+r^{2}_{-}+L^{2})(r_{+}-r_{-})]^{-1}.
\end{equation}
In the continued fraction method, we need to expand the radial function as a power series
\begin{equation}
	\Phi(r)=(r-r_{+})^{N_{+}}(r-r_{-})^{\Delta-N_{+}}\sum_{n=0}^{\infty}a_{n}\left(\frac{r-r_{+}}{r-r_{-}}\right)^{n}.\label{eq:CFM_series}
\end{equation}
Then substituting (\ref{eq:CFM_series}) into (\ref{eq:CFM_KG}), we can obtain a complicated recurrence relation on sequence $\{a_{n}\}$.
Using Gaussian eliminations, the relation about the coefficients can be reduced to the three-term recurrence relation
\begin{equation}
	\begin{aligned}
		0&=C^{0}_{0}a_{0}+C^{-1}_{1}a_{1},\\
		0&=C^{1}_{n}a_{n}+C^{0}_{n+1}a_{n+1}+C^{-1}_{n+2}a_{n+2},\qquad n\geq0
	\end{aligned}\label{eq:CFM_relation}
\end{equation}
where $C^{i}_{n},i=(1,0,-1)$ are the functions of the frequency $w$ and the physical parameters of RN-AdS black hole.
Due to the convergence of the series (\ref{eq:CFM_series}), the continued fraction can be constructed from the relation (\ref{eq:CFM_relation})
\begin{equation}
	0=C^{0}_{0}-\frac{C^{-1}_{1}C^{1}_{0}}{C^{0}_{1}-\frac{C^{-1}_{2}C^{1}_{1}}{C^{0}_{2}-\cdots}}.\label{eq:CFM_fraction}
\end{equation}
At this time, quasi-normal frequencies $w$ are the roots of the polynomial on the right side of the equation (\ref{eq:CFM_fraction}).
The eigenvalue problem is transformed into a problem of finding the roots of the polynomial in the complex plane.
We search for these roots numerically by plotting the logarithm of the continued fraction.

\begin{figure}
	\begin{center}
		\subfigure[]{\includegraphics[width=.49\linewidth]{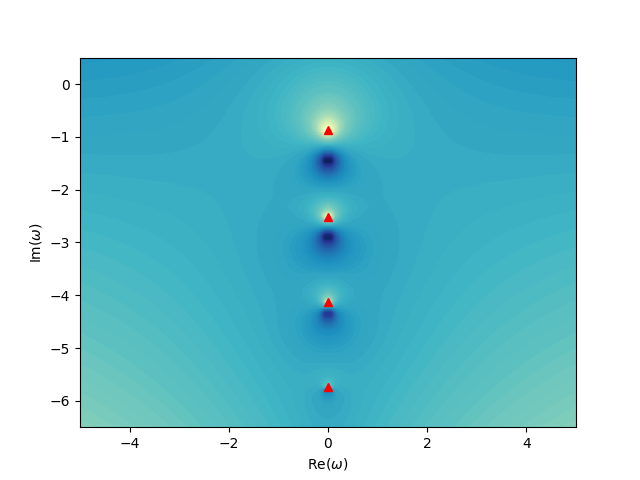}\label{fig:QNM_large}}
		\subfigure[]{\includegraphics[width=.49\linewidth]{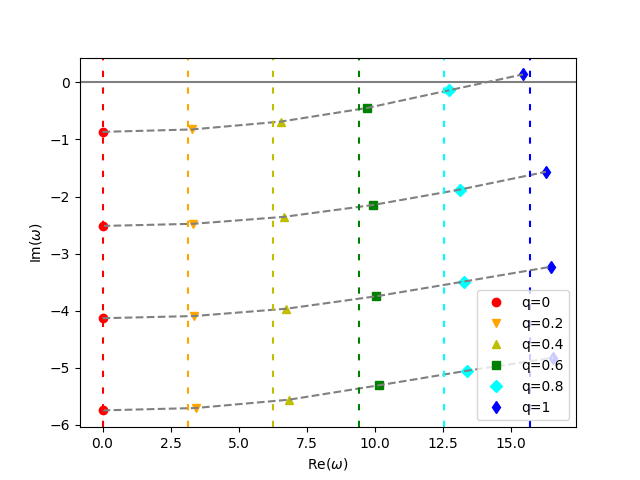}\label{fig:QNM_large_q}}
		\subfigure[]{\includegraphics[width=.49\linewidth]{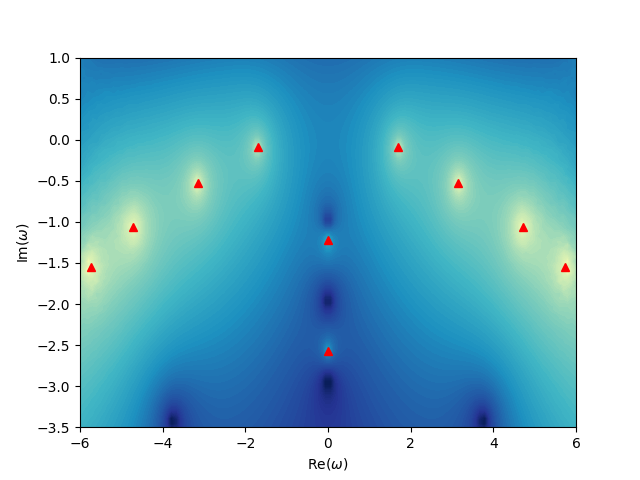}\label{fig:QNM_small}}
		\subfigure[]{\includegraphics[width=.49\linewidth]{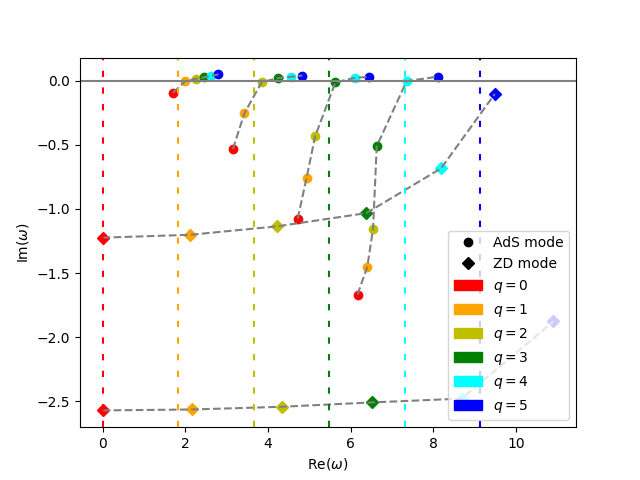}\label{fig:QNM_small_q}}
		\caption{(a, c): The quasi-normal spectrums of large ($r_{+}=5$) and small ($r_{+}=0.1$) RN-AdS black holes with $\alpha=Q/Q_{c}=0.9$ and $q=0$, respectively. The shade of color represents the logarithmic value of the continued fraction. Quasi-normal modes are marked at the red triangle.
			(b, d): The quasi-normal modes of large ($r_{+}=5$) and small ($r_{+}=0.1$) RN-AdS black holes vary with the paratemer $q$, respectively. The vertical dashed lines indicate the superradiance thresholds.}
	\end{center}
\end{figure}

With fixed parameters $\alpha=Q/Q_{c}=0.9$ and $q=0$, the quasi-normal spectrums of large ($r_{+}=5$) and small ($r_{+}=0.1$) RN-AdS black holes are shown in Fig. \ref{fig:QNM_large} and Fig. \ref{fig:QNM_small}, respectively.
For a large RN-AdS black hole, there is a tower of pure imaginary modes extending below the real axis, which is called the zero-damped modes. The zero-damped modes    cluster onto the real axis in the extremal limit \cite{Yang:2012pj,Zimmerman:2015trm,Zimmerman:2016qtn}. 
Based on the discussion in Subsec \ref{subsec:IR}, we expect that the IR instability should appear as the   coupling $q$ increases.
To verify this, we trace the trajectories of these modes as $q$ varies in Fig. \ref{fig:QNM_large_q}.
With increasing $q$, the tower of zero-damped modes shifts to the right along with the superradiance threshold.
The first mode of the zero-damped modes progressively possesses a positive imaginary part, which indicates the occurrence of IR instability.
However, we found the unstable mode is also within the superradiance threshold, which suggests that superradiant instability also works for large RN-AdS black hole.
We have verified other parameters of the large black hole and got the same result: with non-zero $q$, all the unstable IR modes satisfy the superradiance condition (\ref{eq:surcon}).
For small RN-AdS black hole, there is further a branch of AdS modes besides zero-damped modes.
As $q$ increases in Fig. \ref{fig:QNM_small_q}, the AdS modes gradually lag behind the movement of the superradiance threshold and become unstable.
This is exactly the superradiant instability induced by the AdS modes that we expected in the Subsec \ref{subsec:Sup}.
With sufficiently large $q$, the first mode of the zero-damped modes will also become unstable.
But the value of $q$ here is much larger than in the case of large RN-AdS black hole, which shows the IR instability is harder to occur for small RN-AdS black hole.

In conclusion, the dominant unstable modes for large and small RN-AdS black holes come from the zero-damped modes and AdS modes, respectively.
Moreover, they both trigger the superradiant instability since the superradiance condition is alway satisfied when $q\neq 0$.
However, the IR instability prefers large RN-AdS black hole and is suppressed for small RN-AdS black hole.

\section{Nonlinear evolution} \label{sec:nonlinear}
Based on the results of the linear analysis, we further investigate the nonlinear dynamics in this section.
The system of equations (\ref{eq:Einstein_equation}), (\ref{eq:Mx}) and (\ref{eq:KG}) is solved numerically to reveal the processes of scalarization and descalarization, which occur under reflecting boundary conditions and time-dependent source boundary conditions, respectively.
For more details about the numerical procedure, we refer the reader to Appendix \ref{sec:Np}.

In order to reveal the relationships between the evolution of physical quantities and time-dependent sources, we begin with a description of Ward-Takahashi identity in Subsec \ref{sec:WTI}.
Then, we construct the charged hairy black hole according to the two types of instability of RN-AdS in Subsec \ref{sec:Scar}.
Last and foremost, the process of descalarization is realized by opening the source of the scalar field, which is introduced in Subsec \ref{sec:Dscar}.

\subsection{Ward-Takahashi identity}\label{sec:WTI}
From holography, we are required to renormalize the bulk action by adding the following boundary terms \cite{Gibbons:1976ue,Bianchi:2001kw,Elvang:2016tzz}
\begin{equation}
	2\kappa^{2}_{4}S_{\text{reg}}=\int_{M} dx^{4}\sqrt{-g}\mathcal{L}+2\int_{\partial M}dx^{3}\sqrt{-\gamma}K-\int_{\partial M}d^{3}x\sqrt{-\gamma}\left(4+{R}[\gamma]+|\psi|^{2}\right),
\end{equation}
where ${R}[\gamma]$ is the Ricci scalar associated with the boundary induced metric $\gamma_{\mu\nu}$ and $K$ is the trace of extrinsic curvature $K_{\mu\nu}=\gamma^{\sigma}_{\mu}\nabla_{\sigma}n_{\nu}$ with $n_{\nu}$ the outward normal vector field to the boundary.
According to the AdS/CFT correspondence, such a gravity system corresponds to a boundary conformal field theory, where the operators of the fields are defined as

\noindent the scalar operator:
\begin{equation}
	\left\langle O\right\rangle =\kappa_{4}^{2}\lim_{r\rightarrow \infty}\frac{r^{2}}{\sqrt{\gamma}}\frac{\delta S_{\text{ren}}}{\delta{\psi}}=-\frac{1}{2}\lim_{r\rightarrow \infty}r^{2}\left[\psi^{*}+ n^{\mu}(D_{\mu}\psi)^{*}\right],
\end{equation}

\noindent the electric current:
\begin{equation}
	\left\langle J^{i}\right\rangle=\kappa^{2}_{4}\lim_{r\rightarrow \infty}\frac{r^{3}}{\sqrt{\gamma}}\frac{\delta S_{\text{ren}}}{\delta A_{i}}=-\frac{1}{2}\lim_{r\rightarrow \infty}r^{3}n_{\mu}F^{\mu i},
\end{equation}

\noindent the Brown-York tensor:
\begin{equation}
	\left\langle T_{ij}\right\rangle=-2\kappa_{4}^{2}\lim_{r\rightarrow \infty}\frac{r}{\sqrt{\gamma}}\frac{\delta S_{\text{ren}}}{\delta\gamma^{ij}}=\lim_{r\rightarrow \infty}r\left[G[\gamma]_{ij}-K_{ij}-\left(2-K+\frac{1}{2}|\psi|^{2}\right)\gamma_{ij}\right].
\end{equation}
Here $G[\gamma]_{ij}$ is the Einstein tensor associated with $\gamma_{ij}$. Then the variation of the renormalized on-shell action has the form
\begin{equation}
		\kappa^{2}_{4}\delta S_{\text{ren}}=\int_{\partial M}d^{3}x\sqrt{-\gamma_{(0)}} \left(-\frac{1}{2}\left\langle T_{ij}\right\rangle\delta\gamma_{(0)}^{ij}+\left\langle J^{i}\right\rangle\delta A_{i(0)}+\left\langle O\right\rangle\delta\psi_{(0)}+\left\langle O\right\rangle^{*}\delta\psi^{*}_{(0)}\right), \\
\end{equation}
where the subscripts denote the coefficients of the leading order term in the asymptotic behavior at the boundary.
Note that the variation with respect to the bulk fields gives rise to the field equations (\ref{eq:Einstein_equation}), (\ref{eq:Mx}) and (\ref{eq:KG}).

Every kind of invariance preserved by the renormalized on-shell action corresponds to a constraint on the operators of the fields.
One of them is the $U(1)$ gauge invariance 
\begin{equation}
	\delta A_{i}=\nabla_{i}\Theta,\quad \delta \psi=iq\Theta\psi,
\end{equation}
which gives the conservation formula for the electric current
\begin{equation}
	\nabla_{i}\left\langle J^{i}\right\rangle=iq\left(\psi_{(0)}\left\langle O\right\rangle-\psi^{*}_{(0)}\left\langle O\right\rangle^{*}\right).\label{eq:conservation1}
\end{equation}
Another conservation formula is for the energy-momentum tensor
\begin{equation}
	\nabla^{j}\left\langle T_{ij}\right\rangle=F_{ij}\left\langle J^{j}\right\rangle+\left\langle O\right\rangle D_{i}\psi_{(0)}+\left\langle O\right\rangle^{*}(D_{i}\psi_{(0)})^{*},\label{eq:conservation2}
\end{equation} 
which is derived from the diffeomorphism invariance
\begin{equation}
	\delta \gamma^{ij}=\pounds_{\xi}\gamma^{ij},\quad \delta{A}_{i}=\pounds_{\xi}A_{i},\quad \delta\psi=\pounds_{\xi}\psi,
\end{equation}
where $\pounds_{\xi}$ is the Lie derivative with respect to an arbitrary vector field $\xi^{i}$ tangent to the boundary.

With the ansatz (\ref{eq:ansatz}) and the boundary conditions (\ref{eq:asy_behavior}), the $U(1)$ charge and energy of the system have the form
\begin{subequations}
	\begin{align}
		\left\langle J^{t}\right\rangle=&\frac{1}{2}Q,\\
		\left\langle T_{tt}\right\rangle=&2M- \psi_{1} \left\langle O\right\rangle  - \psi^{*}_{1}\left\langle O\right\rangle^{*},
	\end{align}
\end{subequations}
where the symbols $Q$ and $M$ represent electric charge and ADM mass, respectively.
From the Ward-Takahashi identities (\ref{eq:conservation1}) and (\ref{eq:conservation2}), we can obtain the relationships between above physical quantities and the source of the scalar field $\psi_{1}$ in (\ref{eq:asy1}) 
\begin{subequations}
	\begin{align}
		\partial_{t}Q&=2iq\left(\psi_{1}\left\langle O\right\rangle-\psi^{*}_{1}\left\langle O\right\rangle^{*}\right),\\
		\partial_{t}\left\langle T_{tt}\right\rangle&=-\left\langle O\right\rangle D_{t}\psi_{1}-\left\langle O\right\rangle^{*} (D_{t}\psi_{1})^{*}.
	\end{align}\label{eq:QM_WI}
\end{subequations}
When the scalar source disappears, the above quantities are conserved throughout the evolution.


\subsection{Scalarization}\label{sec:Scar}
The nonlinear dynamics of the small and large RN-AdS black holes has been studied in \cite{Bosch:2016vcp} and \cite{Bosch:2019anc}, respectively.
Their final states of evolution share similar properties.
A harmonically oscillating complex scalar field is generated from the unstable RN-AdS black hole background.
The electric charge and mass are extracted from the black hole by the scalar field.
With larger $q$, the scalar condensate is farther away from the black hole and carries more electric charge.
In this subsection, we only construct the hairy black hole in preparation for the descalarization process.
For more details on the dynamics in scalarization, please refer to the above two references. 

\begin{figure}[h]
	\begin{center}
		\subfigure[]{\includegraphics[width=.49\linewidth]{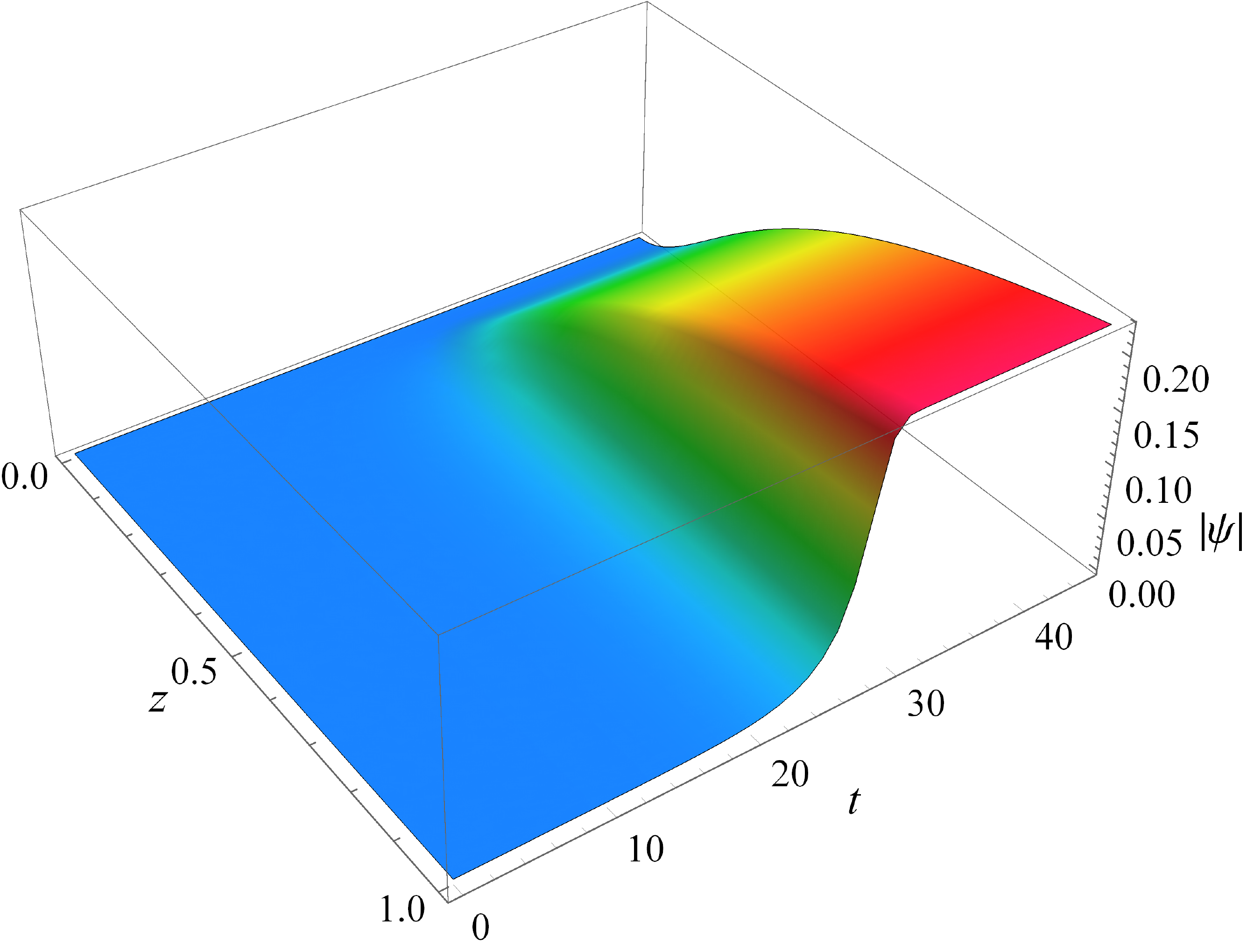}\label{fig:scalarization_large}}
		\subfigure[]{\includegraphics[width=.49\linewidth]{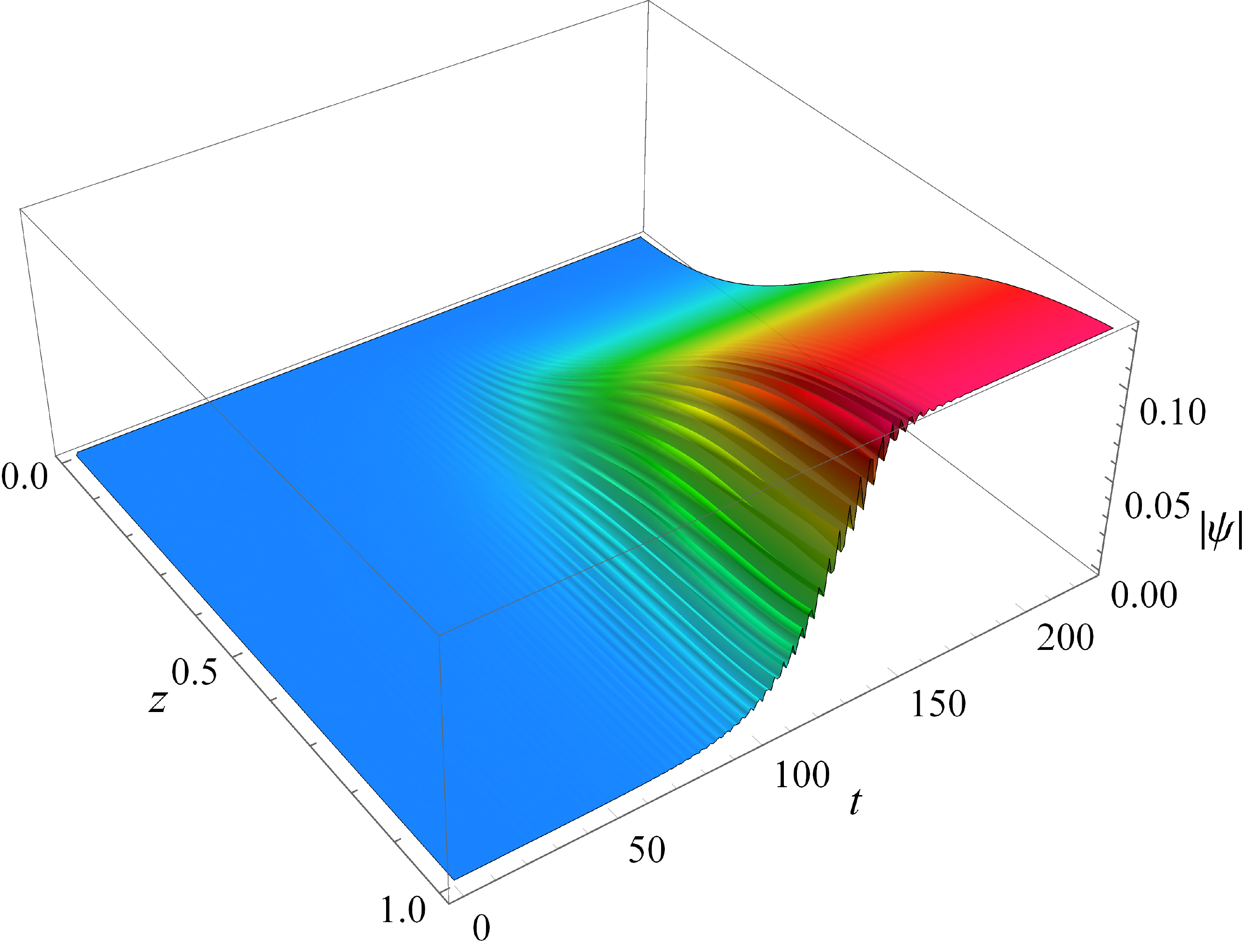}\label{fig:scalarization_small}}
		\caption{The evolution of the configurations of the modulus of the complex scalar fields with the same parameters $\alpha=Q/Q_{c}=0.9$ and $q=8$ during scalarization process. The values of the outer horizon radius of the RN-AdS black holes as the initial data are $r_{+}={1}$ (left), $r_{+}=0.1$ (right), respectively.}\label{fig:scalar_scalarization}
	\end{center}
\end{figure}

Without loss of generality, we take the large ($r_{+}={1}$) 
\footnote{A large RN-AdS black hole generally means that its event horizon radius is larger than the AdS radius. 
Here we refer to the RN-AdS black hole whose dominant unstable mode originates from the zero-damped modes as a large RN-AdS black hole.
In order to highlight the effect of the scalar source on the black hole, we avoid picking situations where the physical parameters (the electric charge and ADM mass) are too large.} 
and small ($r_{+}=0.1$) RN-AdS black holes with the same parameter $\alpha=Q/Q_{c}=0.9$ as the background solutions.
As $q$ increases, the dominant unstable modes of them come from the zero-damped modes and AdS modes, respectively.
For the initial data of scalar field, we choose a Gaussion-like perturbation in the form of $\delta\psi={10^{-4}}z^{2}(1-z)^{2}\text{exp}\left[-50(z-0.5)^{2}\right]$ with a coordinate compactification $z=r^{-1}$ such that the radial direction is bounded in $z\in [0,1]$, where the position of the apparent horizon is fixed at $r_{h}=1$ by the radial shift $r\rightarrow \overline{r}=r+\lambda$ preserved by the ansatz (\ref{eq:ansatz}) during evolution.

The configurations of the modulus of the complex scalar fields as a function of time with $q=8$ are illustrated in Fig. \ref{fig:scalar_scalarization}.
As we see, in both cases, there is a scalar condensation attached to the black hole in the final state of evolution.
However, compared to the case of large black hole, the dynamic process of convergence to the stationary hairy black hole is less smooth for the small RN-AdS black hole case.
During the evolution, the electric charge $Q$ and ADM mass $M$ of the system are conserved due to the vanishing source of scalar field $\psi_{1}$, which is guaranteed by the Ward-Takahashi identities (\ref{eq:QM_WI}).

\subsection{Descalarization}\label{sec:Dscar}
Due to the confining AdS boundary which prohibits matter escaping, the dynamics in this spacetime essentially occur under the microcanonical ensemble as the electric charge and energy of the system are conserved during evolution.
A natural question which arises is how to adjust the asymptotic behaviors of matter fields or metric fields at the AdS boundary to induce changes of the electric charge and energy.
As we can see from the Ward-Takahashi identities (\ref{eq:QM_WI}), they are intrinsically kinetic in the presence of the scalar source.
Therefore, in this subsection, we take the challenge to study the nonlinear dynamics for time-dependent source $\psi_{1}(t)$ on the background solutions of charged hairy black holes.

We take the final states of the scalarization processes obtained in the previous subsection as the initial data of the quench processes.
In the process of quench, we also need to specify a time dependency for the scalar source.
In fact, the scalar source can be seen as an extra parameter in the phase space.
In order to compare the properties of the initial and final states of the quench process under the same value of the scalar source, we choose the following families of quench
\begin{equation}
	\psi_{1}=H\text{exp}\left[-\frac{(t-10)^{2}}{6}\right],\label{eq:quench_G}
\end{equation}
where the parameter $H$ is the strength of the quench and the source decays rapidly to zero at late times.

\begin{figure}[h]
	\begin{center}
		\subfigure[]{\includegraphics[width=.49\linewidth]{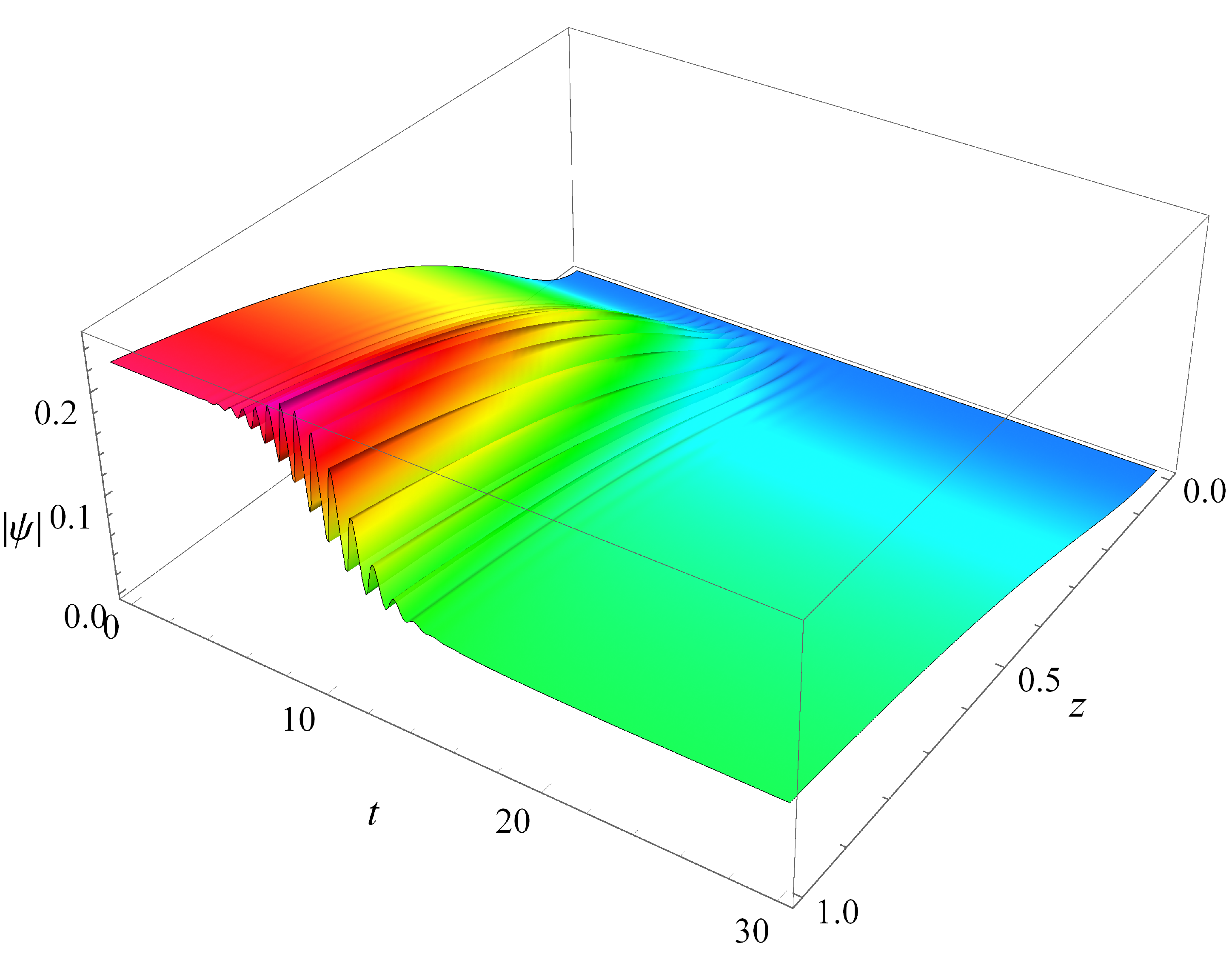}\label{fig:descalarization_1}}
		\subfigure[]{\includegraphics[width=.49\linewidth]{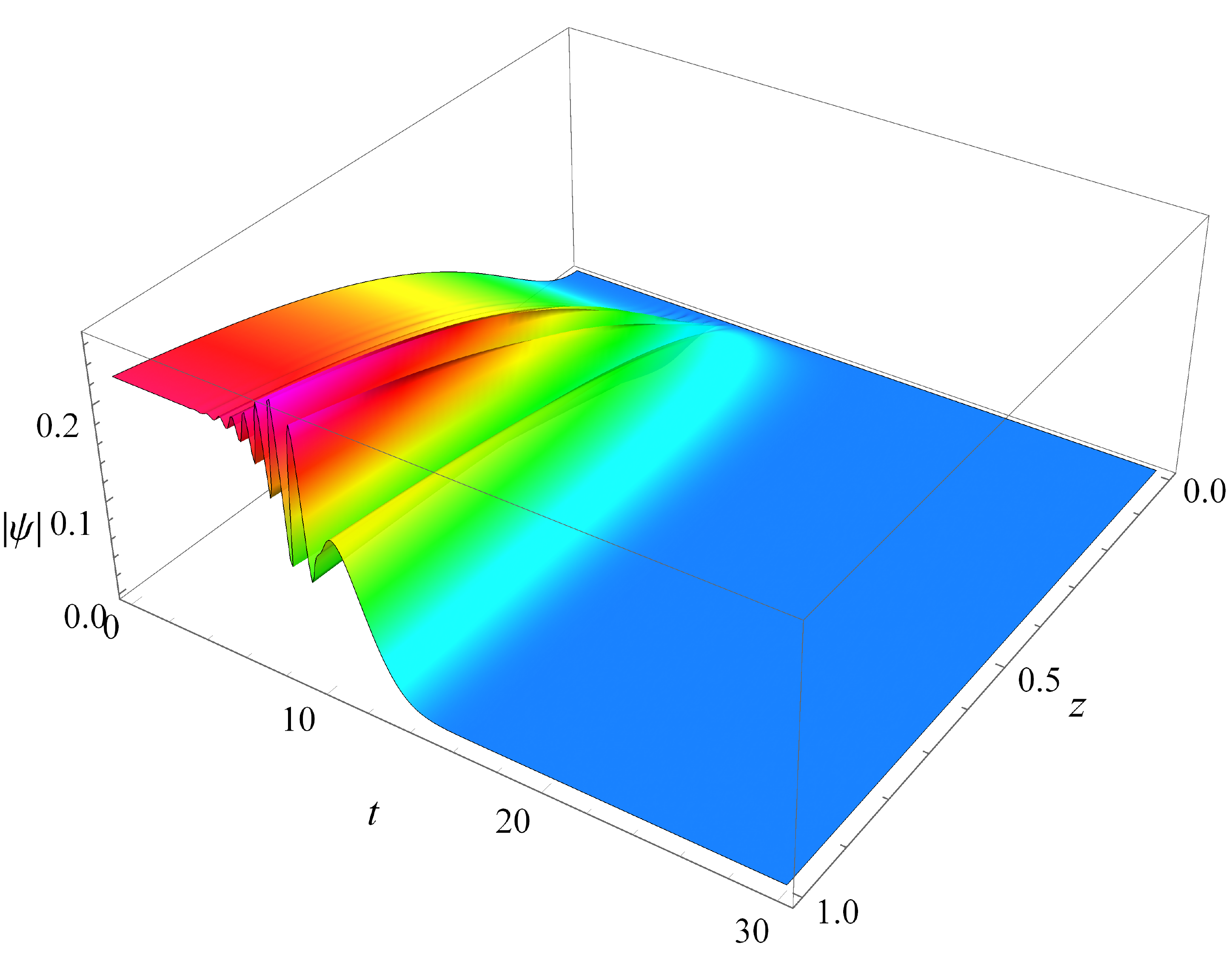}\label{fig:descalarization_2}}
		\caption{The configurations of the modulus of the complex scalar fields as a function of time during the quench process. Both initial data are the final state of Fig. \ref{fig:scalarization_large}, which corresponds to the case of large RN-AdS black hole. The only difference is that the strength of the quench are $H=0.14$ (left) and $H=0.4$ (right), respectively.}\label{fig:scalar_descalarization}
	\end{center}
\end{figure}

To reveal the real-time dynamics of the descalarization process, we impose the quench (\ref{eq:quench_G}) with the strength $H=0.14$ and $H=0.4$ on the final state of Fig. \ref{fig:scalarization_large}, which corresponds to the case of large RN-AdS black hole.
The time dependence of the configuration of the modulus of the complex scalar field is shown in Fig. \ref{fig:scalar_descalarization}.
As we can see from Fig. \ref{fig:descalarization_1}, the scalar field starts to oscillate when the source is present and gradually increases in amplitude with the increase of the source.
Eventually, the system reaches a new equilibrium state as the source decays, which possesses less scalar condensation.
For stronger quench illustrated in Fig. \ref{fig:descalarization_2}, the scalar field oscillates with greater amplitude during quench process and completely subsides when the system reaches the final state, leaving behind a bald black hole.
We have checked that for greater strength than this, the final state of evolution is always a bald black hole, leading to the conclusion that such form of quench (\ref{eq:quench_G}) will cause the charged hairy black hole to be descalaried.
Actually, other types of quench we experimented with, such as rectangular wave, also descalarize the system.

\begin{figure}[h]
	\begin{center}
		\subfigure[]{\includegraphics[width=.49\linewidth]{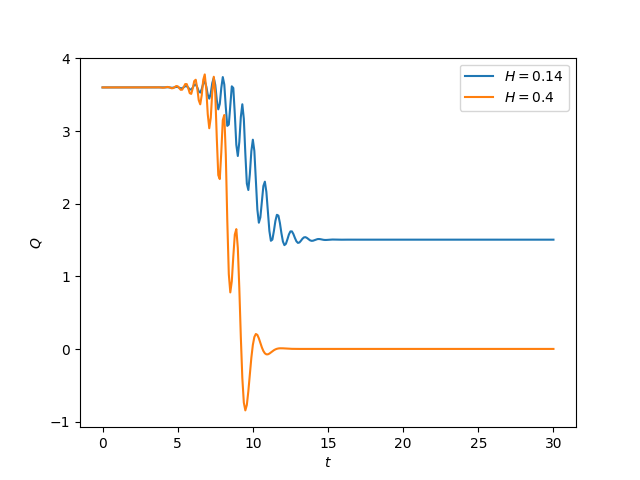}\label{fig:Q_quench}}
		\subfigure[]{\includegraphics[width=.49\linewidth]{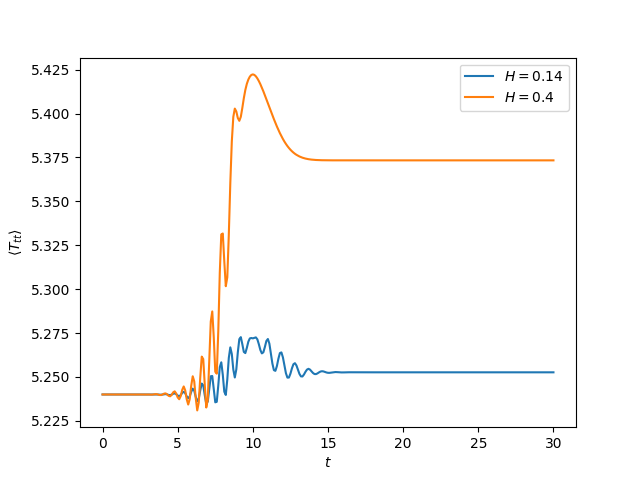}\label{fig:Ttt_quench}}
		\subfigure[]{\includegraphics[width=.49\linewidth]{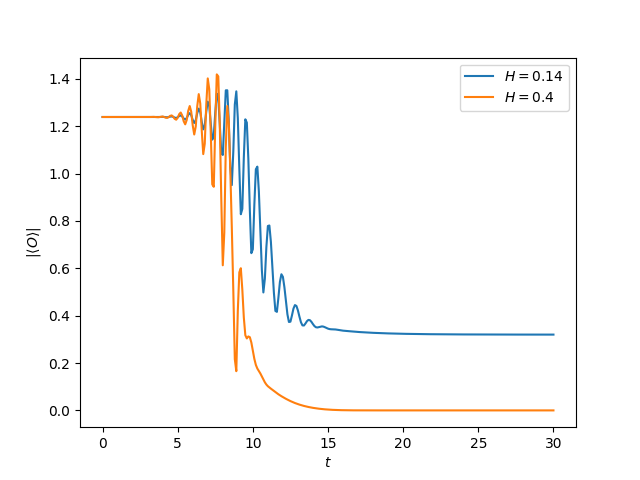}\label{fig:O_quench}}
		\subfigure[]{\includegraphics[width=.49\linewidth]{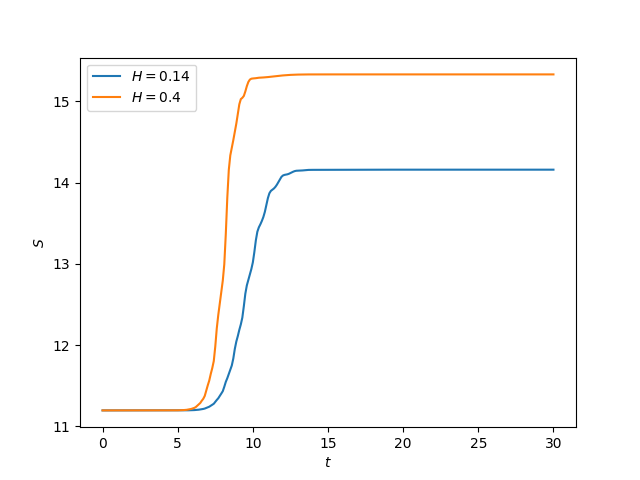}\label{fig:S_quench}}
		\caption{The electric charge (a), energy (b), modulus of scalar operator (c) and entropy (d) of the system as a function of time during the quench process. The entropy is defined as $S=2\pi\Sigma^{2}(r_{h})$, where $\Sigma^{2}(r_{h})$ represents the area of the apparent horizon.  The blue and orange lines represent the cases where the quench strength is $H=0.14$ and $H=0.4$, respectively. }\label{fig:quantity_quench}
	\end{center}
\end{figure}

In the process of quench, in addition to the dynamics of the scalar field, another important result we expect to reveal is the evolution of the physical quantities of the system, which is shown in Fig. \ref{fig:quantity_quench}.
Similar to the behavior of the scalar field, the electric charge (Fig. \ref{fig:Q_quench}) and energy (Fig. \ref{fig:Ttt_quench}) of the system also have the phenomenon of oscillation with the appearance of the scalar source.
And in general they evolve in opposite directions.
Compared with the initial state, the electric charge of the final state is reduced and the energy is increased.
Moreover, the stronger quench brings a larger amount of change to the physical quantities of the initial and final states.
Specially, in the case of $H=0.4$, almost all the electric charge possessed by the system are lost, which indicates the final state of the evolution is a Schwarzschild-AdS black hole.
Fig. \ref{fig:O_quench} once again clearly characterizes the dynamic behavior of the scalar field from the perspective of the scalar operator.
The second law of black hole mechanics requires the entropy of the system never decreases in dynamical processes.
The Fig. \ref{fig:S_quench} indicates that it is satisfied even in the case of temporal dependence of scalar source.

\begin{figure}[h]
	\begin{center}
		\subfigure[]{\includegraphics[width=.49\linewidth]{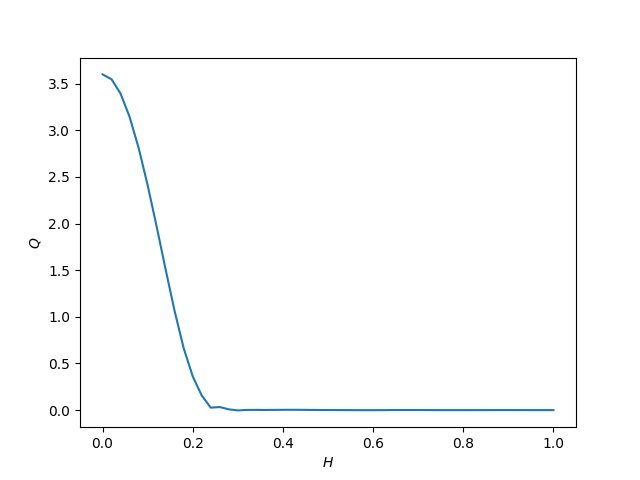}}
		\subfigure[]{\includegraphics[width=.49\linewidth]{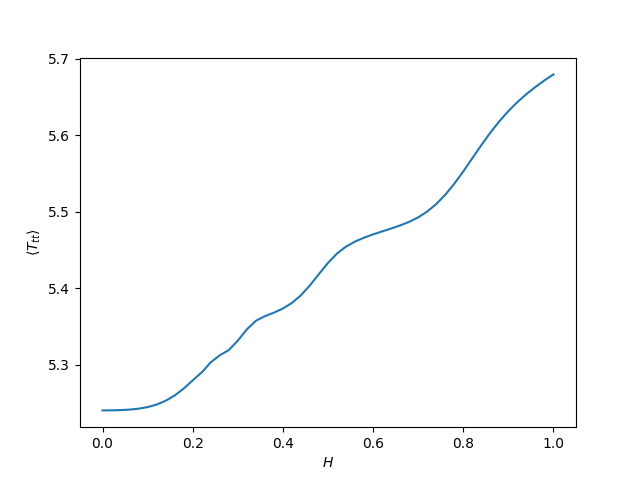}}
		\subfigure[]{\includegraphics[width=.49\linewidth]{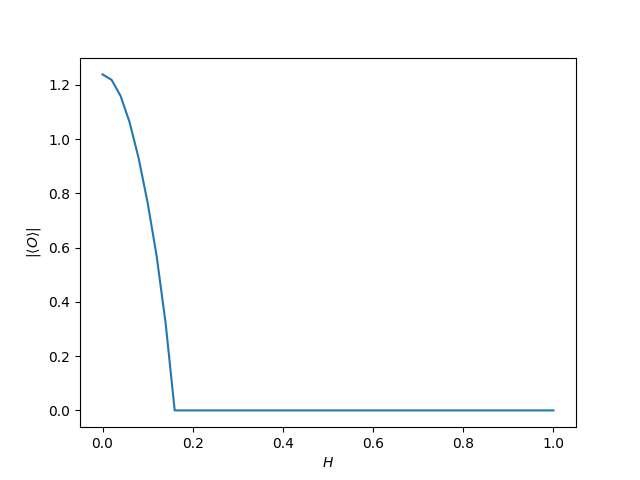}}
		\subfigure[]{\includegraphics[width=.49\linewidth]{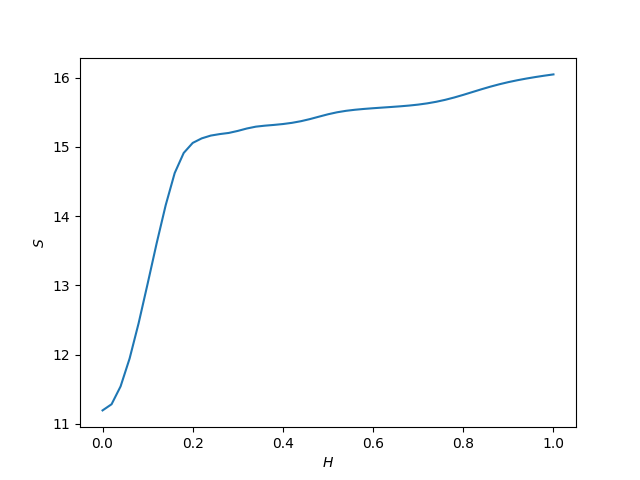}}
		\caption{The variation of electric charge (a), energy (b), modulus of scalar operator (c) and entropy (d) of the final state of the quench process with the quench strength.}\label{fig:quantity_p}
	\end{center}
\end{figure}

\begin{figure}
	\begin{center}
		\subfigure[]{\includegraphics[width=.49\linewidth]{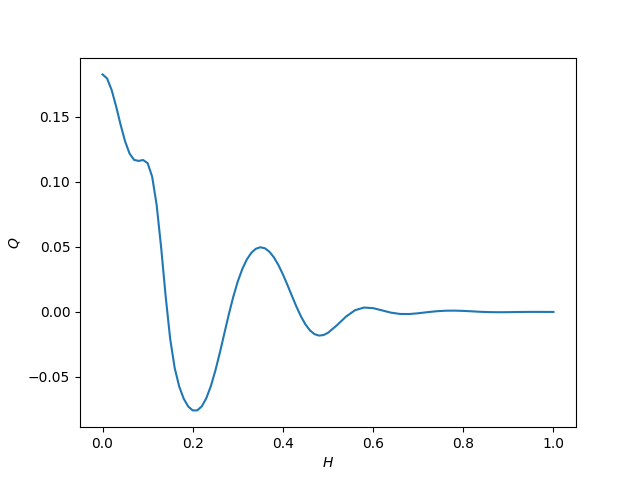}\label{fig:Q_vs_p_small}}
		\subfigure[]{\includegraphics[width=.49\linewidth]{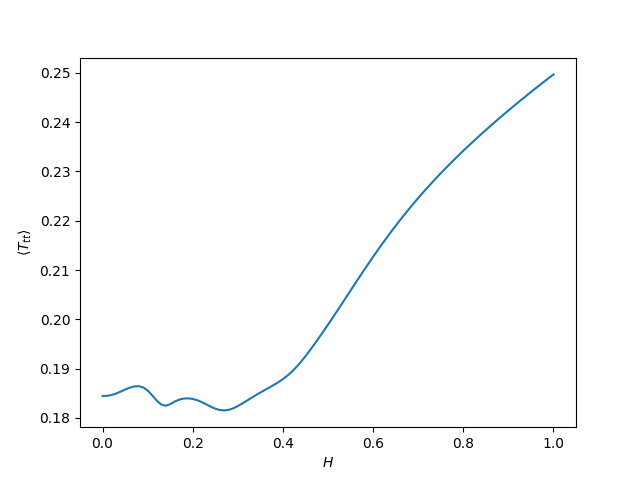}\label{fig:Ttt_vs_p_small}}
		\subfigure[]{\includegraphics[width=.49\linewidth]{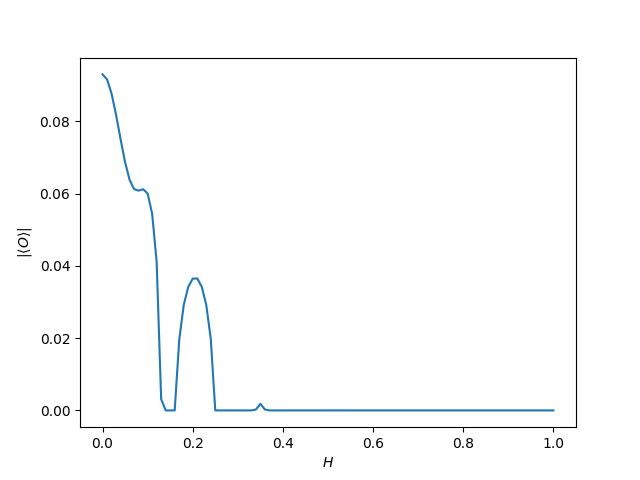}\label{fig:O_vs_p_small}}
		\subfigure[]{\includegraphics[width=.49\linewidth]{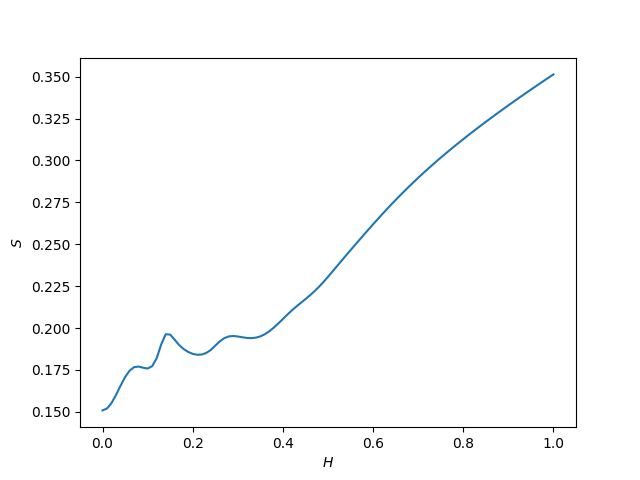}\label{fig:S_vs_p_small}}
		\caption{The electric charge (a), energy (b), scalar operator (c) and entropy (d) of the final state of the quench process as a function of quench strength, where the initial data is the stationary hairy black hole constructed by superradiant instability in Fig. \ref{fig:scalarization_small}.}\label{fig:quench_small}
	\end{center}
\end{figure}

In order to further reveal the effect of quench strength on the system, we draw the variation of physical quantities of the final state of quench process with the quench strength in Fig. \ref{fig:quantity_p}, which can be roughly divided into three parts.
In the region of weak quench strength $(H\lesssim 0.18)$, there is no significant change in the energy of the system.
But the electric charge is drastically reduced, and the scalar operator drops with it.
We refer to this region as the descalarization region where the scalar condensation of the final state is sensitive to the quench strength.
This is also the region where the entropy changes the most.
There is a threshold for quench strength $(H_{O}\approx 0.18)$ where the system is descalarized as a bald black hole.
Note that the electric charge of the system is not completely lost at the threshold, which indicates the final state is a RN-AdS black hole when the quench strength is slightly larger than the threshold.
In the second region $(0.18\lesssim H\lesssim 0.4)$, the electric charge of the system continues to decrease with increasing quench strength until it disappears completely.
Actually, the variation of the electric charge with quench strength is not monotonic, but oscillates and converges to zero, which is more obvious in the case of small RN-AdS black hole shown in Fig. \ref{fig:Q_vs_p_small}.
RN-AdS black hole is the only candidate for this region.
There is also a threshold for quench strength $(H_{Q}\approx 0.4)$ where the system loses almost all electric charge, indicating the final state of evolution is a Schwarzschild-AdS black hole when the quench strength is greater than this threshold. 
In the region of strong quench strength $(H\gtrsim 0.4)$, only the Schwarzschild-AdS black hole is left, because the system has only one degree of freedom, which is energy.
As the quench strength continues to increase, the energy of the Schwarzschild-AdS black hole rises significantly.

For the case of small RN-AdS black hole, there is a similar but different phenomenon.
We still impose the quench (\ref{eq:quench_G}) on the final state of Fig. \ref{fig:scalarization_small} and show the physical quantities of the system vary with the quench strength in Fig. \ref{fig:quench_small}.
As we can see from Fig. \ref{fig:Q_vs_p_small}, the electric charge has an obvious oscillatory behavior with the increase of quench strength, and converges to zero with sufficiently large $H$.
Fig. \ref{fig:Ttt_vs_p_small} indicates we are able to extract energy from AdS space-time in the range of small quench strength.
The intriguing phenomenon shown in Fig. \ref{fig:O_vs_p_small} is that as the strength $H$ increases, the system re-scalarizes again with the negative electric charge after it is fully descalarized.
Eventually, the scalar condensation disappears as the electric charge decays.
At the same time, the  entropy no longer increases monotonically with $H$ when the quench strength is small, which can be seen from Fig. \ref{fig:S_vs_p_small}.

Based on the numerical results, we conclude that the charged hairy black hole can be descalarized during quench process, where the scalar source is time-dependent.
The final state of evolution can be a hairy black hole with less scalar condensation, a RN-AdS black hole or a Schwarzschild-AdS black hole, which depends on the quench strength.
Specially, with sufficient quench strength, the system always loses all the electric charge and evolves to a Schwarzschild-AdS black hole.

\section{Conclusion}\label{sec:Co}
In this paper, we first gave a brief description of the IR instability and superrandiant instability of the RN-AdS black hole, and then calculated the quasi-normal mode spectrum of the RN-AdS black hole using the continued fraction method.
In the cases of unstable modes, we numerically simulated the nonlinear evolution under source-free boundary conditions to obtain the charged hairy black hole.
Finally, we studied the real-time dynamics of the system in the case of temporal dependence of scalar source, where the descalarization is achieved.

At the linear level, the quasi-normal mode spectrums of large and small RN-AdS black holes are studied.
There are two branches of modes: zero-damped modes and AdS modes, which dominate the instabilities of large and small RN-AdS black holes\footnote{That is to say the dominant unstable modes of large and small RN-AdS black holes are generated by zero-damped modes and AdS modes, respectively.}, respectively.
The IR instability is suppressed for the case of small RN-AdS black hole.
Interestingly, all the unstable modes satisfy the superradiance condition when $q\neq 0$ in either case, which indicates the superradiant instability plays a role in both large and small RN-AdS black holes.

At the nonlinear level, these unstable modes grow and drive the system to 
a stationary hairy black hole.
To reveal the response of the hairy black hole to the scalar source, we imposed a time dependent scalar source to the system.
We found that the physical quantities of the system (the electric charge, energy and scalar condensation) start to oscillate with the appearance of the scalar source.
However, the entropy always increases monotonically with time, which is guaranteed by the second law of black hole mechanics.
Eventually, as the scalar source decays, the system reaches a new stable state.
Given a quench of the form ($\ref{eq:quench_G}$), the final state of evolution depends on the quench strength.
For the case of large RN-AdS black hole, as the quench strength increases, the system will evolve to a hairy black hole with less scalar condensation, RN-AdS black hole and Schwarzschild-AdS black hole, respectively.
However, an interesting phenomenon occurs in the case of small RN-AdS black hole.
After the system is descalarized with the increase of quench strength, it may be re-scalarized again, so that the above final states will be repeated.
But in either case, the Schwarzschild-AdS black hole is the ultimate victor with sufficient quench strength.

In most of the previous papers, the dynamics in asymptotically AdS spacetime were studied in the microcanonical ensemble as the electric charge and energy of the system are fixed throughout the evolution.
In this work, we took up the challenge of investigating the fully nonlinear numerical simulations of the charged hairy black hole in an open environment.
And the results reveal a novel descalarization mechanism.
Our work underscores the influence of scalar source on black holes and will shed lights into deep investigations of dynamical mechanisms in some other gravity models.
Especially, it is meaningful to seek a holographic interpretation for the time-dependent scalar source and the descalarization process \cite{Bhaseen:2012gg}.
In addition to the source of scalar field, the response of the black hole to the source of metric field is also worthy of attention.
We leave these works for further research in the future.

\appendix
\section{Numerical procedure}\label{sec:Np}
For the nonlinear evolution, the approach in \cite{Chesler:2013lia}, which has been shown to be amenable to a variety of gravitational dynamics problems in the asymptotically AdS spacetime, is used to solve the coupled field equations numerically.
We adopt the ingoing Eddington-Finkelstein metric ansatz compatible with spherical symmetry
\begin{equation}
	ds^{2}=-2W(t,r)dt^{2}+2dtdr+\Sigma(t,r)^{2}d\Omega^{2}_{2}.\label{eq:ansatz}
\end{equation}
The form of the metric ansatz is preserved by a residual diffeomorphism: arbitrary radial shift  $r\rightarrow \overline{r}=r+\lambda$, which is used to put the apparent horizon at a fixed radial position during evolution.
Since this choice makes the computational domain a fixed interval, the fields can be conveniently discretized with pseudospectral methods.

For the Maxwell field, we take the gauge
\begin{equation}
	A_{\mu}(t,r)dx^{\mu}=A(t,r)dt,
\end{equation}
and the complex scalar field $\psi=\psi(t,r)$.
In order to decouple the Einstein equations, the derivative operator $d_{+}=\partial_{t}+W\partial_{r}$, which is the directional derivative along the outgoing null geodesic, must be introduced.

With these preliminaries in hand, the field equations (\ref{eq:Einstein_equation}), (\ref{eq:Mx}) and (\ref{eq:KG}) take the following simple form.

\noindent Einstein equations:
\begin{subequations}
	\begin{align}
		0=&\Sigma''+\frac{1}{2}|\psi'|^{2}\Sigma,\label{eq:E1}\\
		0=&\left(\Sigma{d_{+}\Sigma}\right)' - \frac{1}{2}\left(3- \frac{1}{4}A'^{2} +|\psi|^{2}\right)\Sigma^{2}- \frac{1}{2  },\label{eq:E2}\\
		0=&W''  +  \frac{2 ({d_{+}\Sigma})'}{\Sigma} - 3 - \frac{1}{4}A'^{2} -|\psi|^{2}+\text{Re} \left[\left({d_{+}\psi} -i q A\psi \right)(\psi')^{*}\right], \label{eq:E3}\\
		0=&{d_{+}d_{+}\Sigma}- W'{d_{+}\Sigma}  + \frac{1}{2}|{d_{+}\psi}- i q A\psi|^{2}\Sigma,\label{eq:E4}
	\end{align}
\end{subequations}

\noindent Maxwell equations:
\begin{subequations}
	\begin{align}
		0&=A''+2A'\frac{\Sigma'}{\Sigma} -2q\text{Im}\left[\psi^{*}\psi'\right],\label{eq:M1}\\
		0&=d_{+}A'+2A'\frac{d_{+}\Sigma}{\Sigma}+2q\text{Im}\left[\psi^{*}\left(d_{+}\psi-iqA\psi\right)\right]\label{eq:M2}
	\end{align}
\end{subequations}

\noindent Scalar equation:
\begin{equation}
	0=\left[\Sigma\left(d_{+}\psi-iqA\psi\right)\right]'+\psi'd_{+}\Sigma+\frac{1}{2}iq A'\Sigma\psi+\Sigma\psi,\label{eq:S}
\end{equation}
where we use primes to denote radial differentiation with respect to $r$.

There is a simple and efficient integration strategy for the set of equations due to the nested structure. 
Once given the data for the scalar field $\psi$ on the time slice $t_{0}$, Eq. (\ref{eq:E1}) is a linear second order ordinary differential equation for the field $\Sigma$, which can be successfully solved with appropriate boundary conditions. 
The next target is the Maxwell field $A$ determined by Eq. (\ref{eq:M1}), whose coefficient and source term depend only on the known values of field $\psi$ and $\Sigma$.
Next, the equations that need to be solved in turn are Eqs. (\ref{eq:E2}), (\ref{eq:S}) and (\ref{eq:E3}), whose solutions give the fields $d_{+}\Sigma$, $d_{+}\psi$ and $W$ respectively.
Since fields $d_{+}\psi$ and $W$ are obtained, the scalar field $\psi$ can be pushed to the next time slice $t_{0}+dt$ by integrating in time for $\partial_{t}\psi$.
The procedure is iterated until the entire simulation is completed.
There are two redundant equations (\ref{eq:E4}) and (\ref{eq:M2}), which are used to detect numerical errors.

All is ready except for the boundary conditions. 
The asymptotic near-boundary behaviors of solutions to field equations take the form
\begin{subequations}
	\begin{align}
		\psi=&\psi_{1}r^{-1}+\psi_{2}r^{-2}+o(r^{-3}),\label{eq:asy1}\\
		\Sigma=&r+\lambda-\frac{1}{4}|\psi_{1}|^{2}r^{-1}+o(r^{-2}),\label{eq:asy2}\\
		A=&\mu-Qr^{-1}+o(r^{-2}),\label{eq:asy3}\\
		d_{+}\psi=&-\frac{1}{2}\psi_{1}+\left(\partial_{t}\psi_{1}-\lambda\psi_{1}-\psi_{2}\right)r^{-1}+o(r^{-2}),\label{eq:asy4}\\
		W=&\frac{1}{2}(r+\lambda)^{2}+\frac{1}{2}-\frac{1}{4}|\psi_{1}|^{2}-\partial_{t}\lambda-Mr^{-1}+o(r^{-2}).\label{eq:asy5}
	\end{align}\label{eq:asy_behavior}
\end{subequations}
The source of the scalar field $\psi_{1}$ is a free parameter with respect to the field equations, which is used to quench the system in this work, and the response of the scalar field $\psi_{2}$ cannot be determined by the near-boundary analysis, whose value depends on the solution throughout the bulk.
The symbols $Q$ and $M$ represent electric charge and ADM mass respectively.
Now, we start to introduce the integral constants for solving the field equations.
First, for Eq. (\ref{eq:E1}), the two integration constants are fixed by the first and second terms in the asymptotic behavior (\ref{eq:asy2}): $\Sigma\sim r+\lambda$.
Second, the coefficients of the asymptotic behavior of the field $A$ (\ref{eq:asy3}) including the function $\mu$ and electric charge $Q$ provide the two integration constants for Eq. (\ref{eq:M1}), where $\mu$ is set to zero by the $U(1)$ gauge freedom.
Then, we use the apparent horizon condition $d_{+}\Sigma(r_{h})=0$ to fix the single integration constant in (\ref{eq:E2}).
Next is Eq. (\ref{eq:S}), whose single integration constant is fixed by the coefficient of the subleading term in (\ref{eq:asy4}), where $\psi_{2}$ can be extracted from the asymptotic behavior of the scalar field (\ref{eq:asy1}).
Finally, for Eq. (\ref{eq:E3}), we choose the gauge such that the position of the apparent horizon is time invariant, which gives a boundary condition for $W$ at the horizon from (\ref{eq:E4})
\begin{equation}
	W(r_{h})= - \left.\frac{|{d_{+}\psi}- i q A\psi|^{2}\Sigma}{2(d_{+}\Sigma)'}\right|_{r_{h}}.\label{eq:W_h}
\end{equation}
Besides the boundary condition (\ref{eq:W_h}) fixing one of the two integration constants, another integration constant is fixed by the leading term in the asymptotic behavior (\ref{eq:asy5}).

\section*{Acknowledgement}
This research is partly supported by the Natural Science Foundation of China (NNSFC)
under Grant Nos. 11975235, 12005077, 12035016, 12075202 and Guangdong Basic and Applied Basic Research
Foundation under Grant No. 2021A1515012374.
\bibliographystyle{unsrt}
\bibliography{references}
\end{document}